\input{aipcheck}

\documentclass[
    ,final            % use final for the camera ready runs
  ]
  {aipproc}

\layoutstyle{6x9}

\begin{document}

\title{Ab-initio Green's Functions Calculations of Atoms}


\author{C. Barbieri}{
  address={Theoretical Nuclear Physics Laboratory, RIKEN Nishina Center,
         2-1 Hirosawa, Wako, Saitama 351-0198 Japan}
}

\author{D. Van Neck}{
  address={Center for Molecular Modeling, Ghent University, 
           Proeftuinstraat 86, B-9000 Gent, Belgium}
}

%\author{W. H. Dickhoff}{
%  address={Department of Physics, Washington University,
%           St. Louis, MO 63130, USA}
%}

\begin{abstract}
 The Faddeev random phase approximation (FRPA) method is applied to
calculate the ground state and ionization energies of simple atoms.
First ionization energies agree with the experiment at the level
of $\sim$10~mH or less. Calculations with similar accuracy are expected
to provide information required for developing the proposed quasiparticle-DFT
method.

\end{abstract}

\maketitle

%%%%%%%%%%%%%%%%%%%%%%%%%%%%%%%%%%%%%%%%%%%%
%% MAINMATTER
%%%%%%%%%%%%%%%%%%%%%%%%%%%%%%%%%%%%%%%%%%%%

\paragraph{Introduction}
{\em Ab initio} treatments of electronic systems become unworkable for 
sufficiently complex systems.
On the other hand, the Kohn-Sham formulation~\cite{Koh.65} of density 
functional theory (DFT)~\cite{Hoh.64} incorporates many-body correlations 
(beyond Hartree-Fock), while only single-particle (sp) equations 
must be solved.
 Due to this simplicity DFT is the only feasible approach in some modern 
applications of electronic structure theory. There is therefore a continuing 
interest in studying conceptual improvements and extensions to the DFT
framework.
%In particular it 
%is found that DFT can handle short-range interelectronic correlations quite 
%well, while there is room for improvements in the description of long-range 
%(van der Waals) forces and dissociation processes.%~\cite{aa}.

 An approach in this direction has been proposed in Ref.~\cite{Van.06}
by developing a quasi-particle (QP)-DFT formalism. In the QP-DFT the full
spectral  function is decomposed in the contribution of the QP excitations, 
and a remainder or background part. It is sufficient to have a 
functional model for the energy-averaged background part to set up a 
single-electron self-consistency problem that generates the QP excitations.   
Such an approach is appealing since it contains the well-developed 
standard Kohn-Sham formulation of DFT as a special case, while  at 
the same time emphasis is put on the correct description of QPs, in the 
Landau-Migdal sense~\cite{Mig.67}. Hence, it can provide an improved 
description of the dynamics at the Fermi surface.  Given the close relation 
between QP-DFT and the Green's functions (GF) formulation of many-body 
theory~\cite{FetWal,DicVan},  it is natural to employ {\em ab initio} 
calculations in the latter formalism to investigate the structure 
of possible QP-DFT functionals. 
%
%In this respect it is imperative to identify which classes of 
%diagrams are responsible for the correct description of the QP physics. 
%
This talk reports on recent advances in such calculations.

\paragraph{The Faddeev-RPA  (FRPA) method}

\begin{figure}
  \hbox{
  \includegraphics[height=.21\textheight]{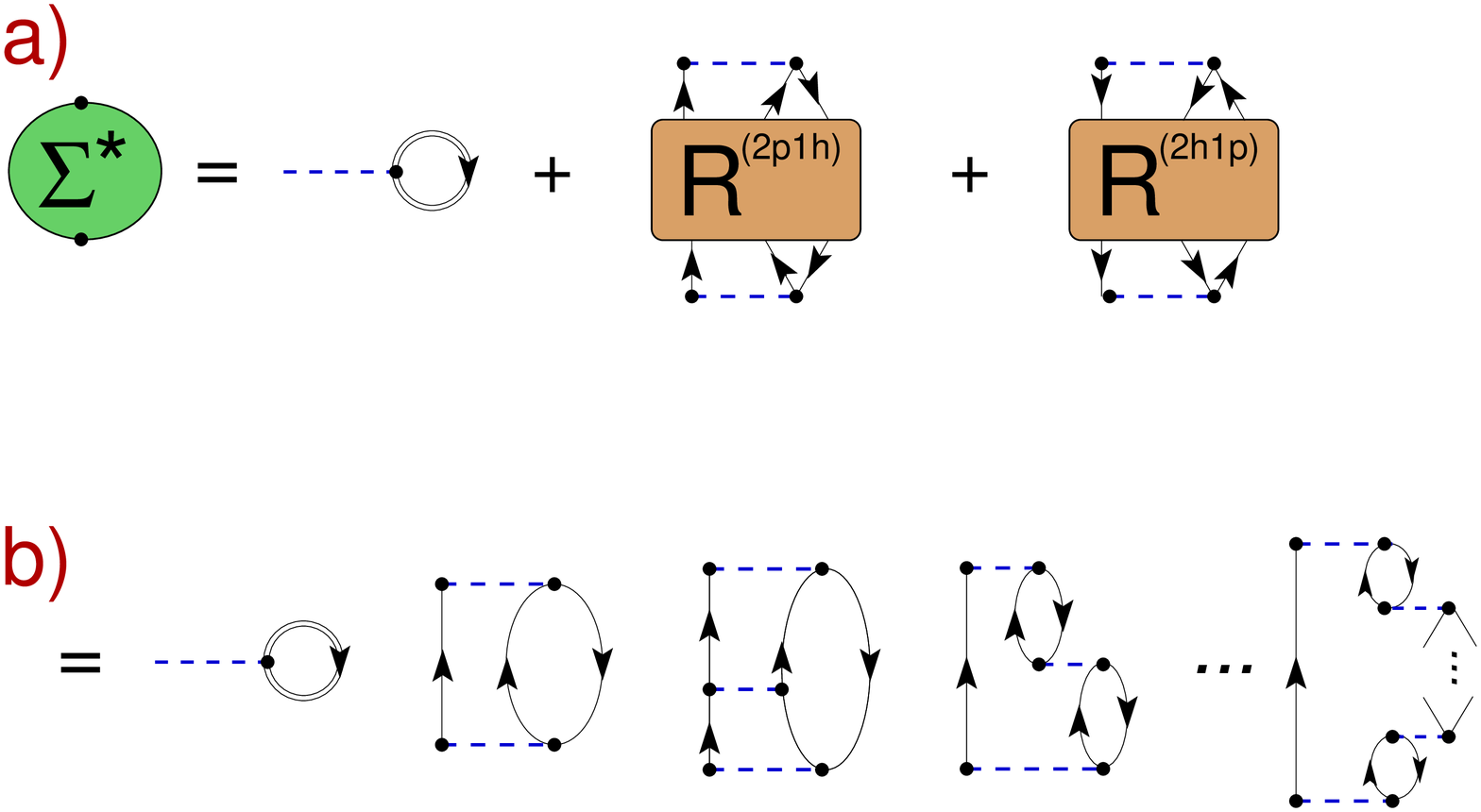}
  \caption{The self-energy $\Sigma^\star(\omega)$ separates exactly into
    a mean field term and the polarization propagators $R(\omega)$ for the
    2p1h/2h1p motion, as shown in a). The double line represent the correlated
    propagator of Eq.~(\ref{eq:g1}).
     Upon expansion of $R(\omega)$ in Feynman diagrams, one obtains the 
    series of diagrams b) for the self-energy.
     The diagram c) gives an example of the contributions to $R^{2p1h}(\omega)$
    that are summed to all orders by the FRPA method.
    \label{frpa_diags} }
  }
  \hspace{0.4cm}
  \hbox{
  \includegraphics[height=.19\textheight]{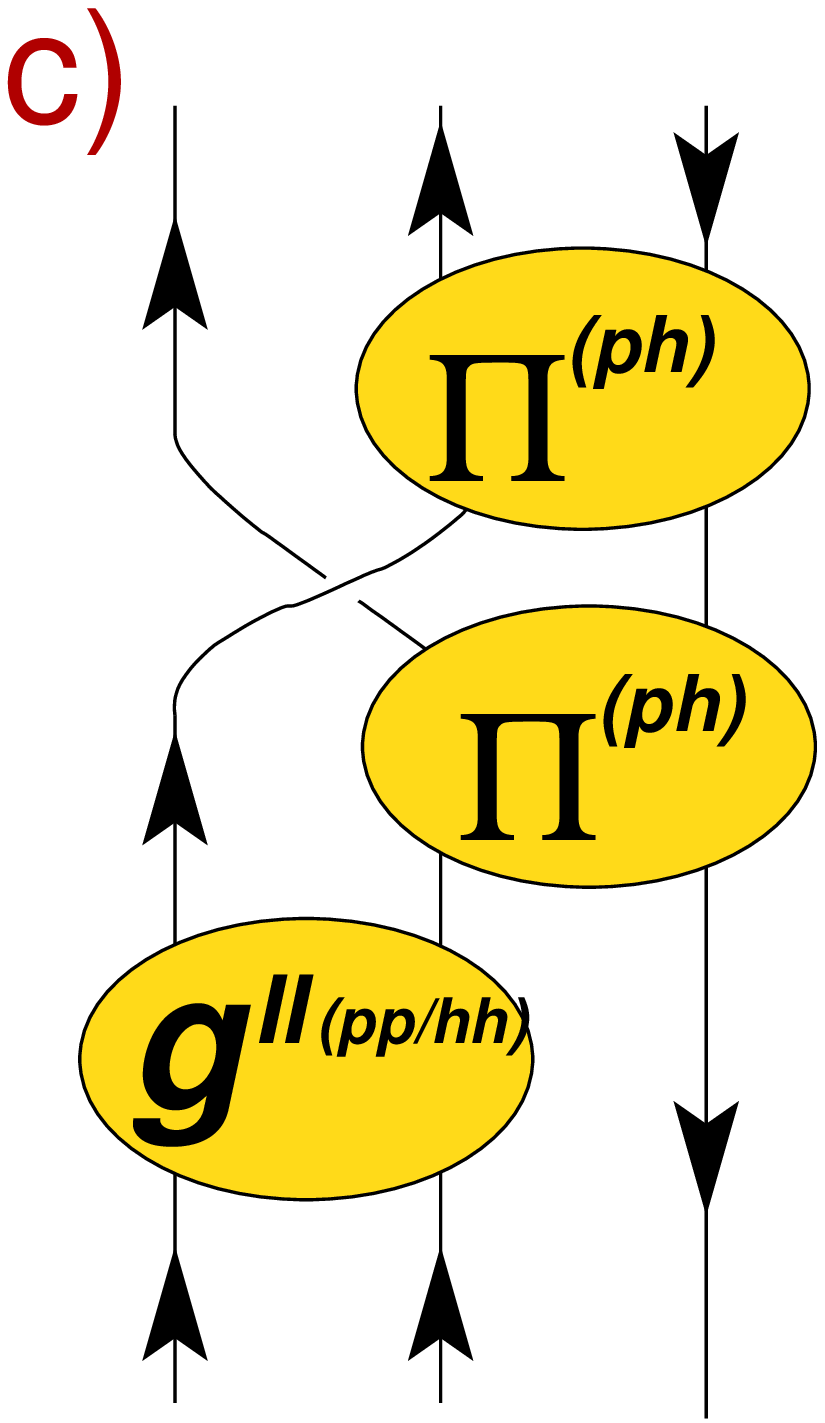}
  \spaceforfigure{.5cm}{1.cm}
  }
\end{figure}

The theoretical framework of the present study is that of propagator theory, 
where the object of interest
is the sp propagator~\cite{FetWal,DicVan},
\begin{equation}
 g_{\alpha \beta}(\omega) ~=~ 
 \sum_n  \frac{ 
          \langle {\Psi^N_0}     \vert c_\alpha        \vert {\Psi^{N+1}_n} \rangle
          \langle {\Psi^{N+1}_n} \vert c^{\dag}_\beta  \vert {\Psi^N_0} \rangle
              }{\omega - (E^{N+1}_n - E^N_0) + i \eta }  ~+~
 \sum_k \frac{
          \langle {\Psi^N_0}     \vert c^{\dag}_\beta  \vert {\Psi^{N-1}_k} \rangle
          \langle {\Psi^{N-1}_k} \vert c_\alpha        \vert {\Psi^N_0} \rangle
             }{\omega - (E^N_0 - E^{N-1}_k) - i \eta } \; ,
\label{eq:g1}
\end{equation}
where $\alpha$,$\beta$,..., label a complete orthonormal basis set and
$c_\alpha$~($c^\dag_\beta$) are the corresponding second quantization 
destruction (creation) operators.
In these definitions, $\vert\Psi^{N+1}_n\rangle$, $\vert\Psi^{N-1}_k\rangle$ 
are the eigenstates, and $E^{N+1}_n$, $E^{N-1}_k$ the eigenenergies of the 
($N\pm1$)-electron system. Therefore, the poles of the propagator reflect the 
electron affinities and ionization energies. For a two-body hamiltonian,
Eq.~(\ref{eq:g1}) also yields the total binding energy via the
Migdal-Galitski\u{\i}-Koltun sum rule~\cite{DicVan}.
The one-body Green's function is computed by solving the Dyson equation
\begin{equation}
 g_{\alpha \beta}(\omega) =  g^{0}_{\alpha \beta}(\omega) \; +  \;
   \sum_{\gamma \delta}  g^{0}_{\alpha \gamma}(\omega) 
     \Sigma^\star_{\gamma \delta}(\omega)   g_{\delta \beta}(\omega) \; \; ,
\label{eq:Dys}
\end{equation}
where the irreducible self-energy $\Sigma^\star_{\gamma \delta}(\omega)$ acts
as an effective, energy-dependent, potential. The latter can be expressed in
terms of the exact propagator $g_{\alpha \beta}(\omega)$, which itself
is a solution of Eq.~(\ref{eq:Dys}), and the polarization propagator,
$R(\omega)$, that accounts for deviations from the mean-field~\cite{DiB.04}. 
This is shown in Fig.~\ref{frpa_diags}a in terms of Feynman diagrams.
The polarization propagator $R(\omega)$ is also expanded in terms of simpler
propagators that involve the propagation of one quasiparticle
[Eq.~(\ref{eq:g1})] or more. This approach has the advantage to help identifing
key physics ingredients of the many-body dynamics. By truncating to particular
subsets of diagrams, one can then construct suitable approximations to
the self-energy. Moreover, since infinite sets of linked diagrams are summed
the approach is non-perturbative and satisfies the extensivity condition. This
expansion also serves as guideline for systematic improvements of the method.

 In the following we are interested in describing the coupling of
sp motion to particle-hole (ph) and two-particle (pp) or two-hole (hh) collective
excitations of the system.
Following Ref.~\cite{Bar.01}, we first calculate the corresponding propagators
by solving the random phase approximation (DRPA) equations in the ph and pp/hh
channels. These are then inserted in the self-energy by solving a set of
Faddeev equations for the $R^{(2p1h)}(\omega)$ and $R^{(2h1p)}(\omega)$
propagators. Fig.~\ref{frpa_diags}c gives an example of one of the diagrams
generated by this procedure.

The details of the
Faddeev RPA (FRPA) approach are given in Refs.~\cite{Bar.01,Bar.07}.
For the present discussion it is sufficient to note that including only ph
propagators corresponds to the same physics of the $GW$~\cite{Hed.65} approach.
This is known to give accurate binding energies for the electron gas, where
the RPA is required to screen the long range Coulomb force. The FRPA
method goes beyond the $GW$ since it accounts completely for Pauli
correlations at the 2p1h/2h1p level and include the propagation of pp/hh
configurations. The latter give crucial contribution to ionization energies
in small systems~\cite{Wal.81}.

\paragraph{Results}
FRPA calculations were performed using the correlation consistent cc-pVTZ and cc-pVQZ
gaussian bases for all atoms except for the ground state of Mg. For the latter
the core-valence version cc-pCV(TQ)Z were used, which include additional compact
gaussians to improve the description of the core electrons. 
This choice was seen to improve the convergence of the binding energy for this atom.
 The results were then extrapolated to the basis set limit according to
\begin{equation}
E_X = E_\infty + A X^{-3} \; , 
\label{extrap}
\end{equation}
where $X=T,Q$ is the cardinal number of the basis. This relation is known to 
give proper extrapolations for correlation energies. Here we apply
it to ionization energies as well, remembering that these are also differences
between eigenenergies.  For the case of Ne, this agrees well with the
ionization energies obtained in a larger basis~\cite{Bar.07} (see below) and
gives us confidence on the extrapolation procedure. 
We estimate that results given in Tabs.~\ref{tab_gse} and~\ref{tab_ie} are
accurate within a few~mH.

\begin{table}
\begin{tabular}{rcccccc}
\hline
  &  & \tablehead{1}{c}{b}{Hartree-Fock}
  &  & \tablehead{1}{c}{b}{FRPA}
  &  & \tablehead{1}{c}{b}{Experiment~\cite{NIST,Tho.01}}   \\
\hline
He  & &    -2.860 (+44)   &  &   -2.903 (+1)   & &   -2.904 \\
Be  & &   -14.573 (+94)   &  &  -14.643 (+24)  & &  -14.667 \\
Ne  & &  -128.547 (+281)  &  & -128.917 (+11)  & & -128.928 \\
Mg  & &  -199.617 (+426)  &  & -200.058 (-15)  & & -200.043 \\
\hline
\end{tabular}
\caption{Hartree-Fock and Faddeev-RPA binding energies (in Hartree) extrapolated from
  the cc-pVTZ and cc-pVQZ basis sets. The deviations from the experiment are
  indicated in parentheses (in mH). For Mg, the cc-pCV(TQ)Z bases were used.}
\label{tab_gse}
\end{table}

Table~\ref{tab_gse} shows the results for the FRPA ground state energies and
compares them to the experiment and the corresponding Hartree-Fock results.
FRPA gives practically exact results for the two electron problem (He) while
it accounts for 96\% of the correlation energy in the larger systems.
 The atom of Be is an exception to this trend, due to the fact that this is
not a good closed shell system. This leads to very soft excitations in
the J$^\pi$=1$^-$,S=1 channel which can drive the ph RPA equation
to instability. A proper treatment of this system may require improving
the treatment of the excitation spectrum beyond the RPA.

\begin{table}[t]
\begin{tabular}{rcccccccc}
\hline
  & & \tablehead{1}{c}{b}{Hartree-Fock}
  & & \tablehead{1}{c}{b}{2$^{nd}$  order}
  & & \tablehead{1}{c}{b}{FRPA}
  & & \tablehead{1}{c}{b}{Experiment~\cite{NIST,Tho.01}}   \\
\hline
He: 1s &  &  0.918 (+14)   &  &  0.906 (+2)   &  &  0.900 (-4)   &  &  0.904 \\ \\
Be: 2s &  &  0.309 (-34)   &  &  0.320 (-23)  &  &  0.322 (-21)  &  &  0.343 \\
    1s &  &  4.733 (+200)  &  &  4.620 (+87)  &  &  4.540 (+7)   &  &  4.533 \\ \\
Ne: 2p &  &  0.850 (+57)   &  &  0.763 (-30)  &  &  0.803 (+10)  &  &  0.793 \\
    1s &  &  1.931 (+149)  &  &  1.750 (-32)  &  &  1.795 (+13)  &  &  1.782 \\ \\
Mg: 3s &  &  0.253 (-28)   &  &  0.274 (-7)   &  &  0.277 (-4)   &  &  0.281 \\
    2p &  &  2.281 (+161)  &  &  2.146 (+26)  &  &  2.130 (+10)  &  &  2.12  \\ \\
Ar: 3p &  &  0.590 (+11)   &  &  0.585 (+6)   &  &  0.578 (-1)   &  &  0.579 \\
    3s &  &  1.276 (+201)  &  &  1.159 (+84)  &  &  1.065 (-10)  &  &  1.075 \\
    2p &  &  9.570 (+410)  &  &  9.519 (+359) &  &  9.199 (+39)  &  &  9.160 \\
\hline
\end{tabular}
\caption{Ionization energies obtained in Hartree-Fock, in second order perturbation
  theory for the self energy and with the full Faddeev-RPA (in Hartree). All results
  are extrapolated from the cc-pVTZ and cc-pVQZ basis sets. The deviations from
  the experiment are indicated in parentheses (in mH). }
\label{tab_ie}
\end{table}

Ionization energies are shown in Tab.~\ref{tab_ie}.
The extrapolated results deviate from experiment by about 5~mH for the first ionization
energies, while it increases to 10-15~mH for the separation of slightly deeper electron 
orbits.
The table also reports the predictions from Hartree-Fock theory and the second order
self-energy (obtained by retaining just the first two diagrams of Fig.~\ref{frpa_diags}b).
Second order corrections account for a large part of correlations but still
lead to sizable errors. The additional correlations included in the present calculations
appear to reduce this error substantially, in particular for deeper electron orbits.
The importance of a treatment that is consistent with at least third order perturbation
theory was already pointed out by Schirmer and co-workers in Ref.~\cite{Wal.81}. Such
contribution were then included in the algebraic diagrammatic construction method
at third order [ADC(3)]~\cite{Sch.83}.
The present formulation of the FRPA includes the ADC(3) completely and gives similar
results for small atomic systems. At the same time, the explicit inclusion of RPA 
phonons holds the promise for successful applications to extended systems. Further work
will be required to verify that this is indeed the case.

\begin{figure}[t]
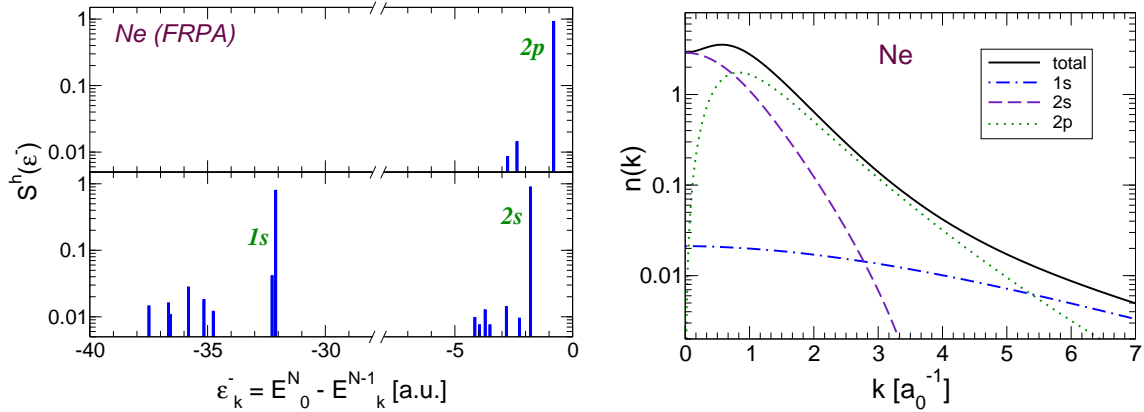

  \hbox{
  \includegraphics[height=.37\textwidth]{fig2a}
  }
  \hspace{0.1cm}
  \hbox{
  \includegraphics[height=.36\textwidth]{fig2b}
  \spaceforfigure{.5cm}{1.cm}
  }
  \caption{Hole spectral function (left) and momentum distribution (right)
    of the Ne atom. The dotted, dashed and dot-dashed lines are the 
    contributions coming from the main $2p$, $2s$ and $1s$ quasihole
    peaks seen on the left side.
 \label{Ne_dist} }
\end{figure}

The Ne atom was also computed in the FRPA approach by using a Hartree-Fock basis
with a discretized continuum~\cite{Bar.07}. The size of the basis set was chosen
large enough to approach the basis set limit. This gave first and second
ionization energies of 0.801 and 1.795~H, in good agreement
with Tab.~\ref{tab_ie}.
 The total binding energy obtained is 128.888~H, somewhat in disagreement with
the extrapolation from the cc-pV(TQ)Z bases and the experiment.
The hole spectral function and momentum distribution of Ne are shown
in Fig.~\ref{Ne_dist}. A moderate tail is generated at large momenta and
originates from the $1s$ orbit at $\approx$-32~H. This is reproduced already
at the Hartree-Fock level and it is due to the fact that the core electrons 
have the highest velocity in a $1/r$ potential. 
We observe that this is a different situation to that of atomic nuclei, where
high momentum components are seen at very large separation energies~\cite{Roh.04}.
 In the latter case the relevant strength is not found at fixed energies
but distributed along a ridge in the energy-momentum plane with $E \approx k^2/2m$,
which signals the presence of strong two-body correlations~\cite{DiB.04}.

\paragraph{Conclusions} The Faddeev RPA method is an expansion of the 
many-body self-energy that makes explicit the coupling between particle and
collective phonons. First applications to small atoms have been reported in
this talk and error for ionization energies were found to be of $\sim$10~mH.
Accurate calculations of quasiparticle properties will serve in developments
of the proposed QP-DFT~\cite{Van.06}.
Due to the inclusion of RPA, it is expected that the present FRPA formalism
could reach similar accuracy also for extended systems. This will be the 
topic of future research efforts.

%%%%%%%%%%%%%%%%%%%%%%%%%%%%%%%%%%%%%%%%%%%%%%%%
%% BACKMATTER
%%%%%%%%%%%%%%%%%%%%%%%%%%%%%%%%%%%%%%%%%%%%%%%%

\begin{theacknowledgments}
  We acknowledge several useful discussions with Prof. W. H. Dickhoff.
\end{theacknowledgments}



\bibliographystyle{aipproc}   % if natbib is available

%\bibliography{sample}

\IfFileExists{\jobname.bbl}{}
 {\typeout{}
  \typeout{******************************************}
  \typeout{** Please run "bibtex \jobname" to optain}
  \typeout{** the bibliography and then re-run LaTeX}
  \typeout{** twice to fix the references!}
  \typeout{******************************************}
  \typeout{}
 }

%%%%%%%%%%%%%%%%%%%%%%%%%%%%%%%%%%%%%%%%%%%
%% The following lines show an example how to produce a bibliography
%% without the help of the BibTeX program. This could be used instead
%% of the above.
%%%%%%%%%%%%%%%%%%%%%%%%%%%%%%%%%%%%%%%%%%%

\end{document}